\documentclass[aps,preprint,showpacs,superscriptaddress,groupedaddress]{revtex4}  % for double-spaced preprint
\usepackage{array,multirow,graphicx}  % needed for figures
\usepackage{amsmath}
\usepackage{amssymb}   % for math
\usepackage{color}
\usepackage{hyperref}
\usepackage{ulem}
\usepackage{natbib}
\usepackage{enumitem}

\newcommand{\ltsima} {$\; \buildrel < \over \sim \;$}
\newcommand{\gtsima} {$\; \buildrel > \over \sim \;$}
\newcommand{\lta} {\lower.5ex\hbox{\ltsima}}
\newcommand{\gta} {\lower.5ex\hbox{\gtsima}}

\def\ln{\mathrm{ln}}

\newcommand{\A}{{\sf{A}}}

\newcommand{\vtheta}{{\boldsymbol{\theta}}}
\newcommand{\vmu}{{\boldsymbol{\mu}}}
\newcommand{\post}{\tilde p(\vtheta|\vx,M)}

\providecommand{\Planck}{\textit{Planck}}

\providecommand{\text}[1]{\rm{#1}}

\newcommand{\begm}{\begin{pmatrix}}
\newcommand{\enm}{\end{pmatrix}}

\newcommand\be{\begin{equation}}
\newcommand\ee{\end{equation}}

\newcommand{\boldvec}[1]{{\mbox{\boldmath{$#1$}}}}

\newcommand{\vD}{\boldvec{D}}

\newcommand{\vx}{\boldvec{x}}

\def\pmb#1{\setbox0=\hbox{#1}%
    \kern-.025em\copy0\kern-\wd0
    \kern.05em\copy0\kern-\wd0
    \kern-.025em\raise.0433em\box0}
\def\ltsima{$\; \buildrel < \over \sim \;$}
\def\gtsima{$\; \buildrel > \over \sim \;$}
\def\simlt{\lower.5ex\hbox{\ltsima}}
\def\simgt{\lower.5ex\hbox{\gtsima}}

\begin{document}

\title{Marginal Likelihoods from Monte Carlo Markov Chains}

\author{Alan Heavens}\email{a.heavens@imperial.ac.uk}
\affiliation{Imperial Centre for Inference and Cosmology (ICIC), Imperial College, Blackett Laboratory, Prince Consort Road, London SW7 2AZ, U.K.}
\author{Yabebal Fantaye}
\affiliation{African Institute for Mathematical Sciences,
6–8 Melrose Road,
Muizenberg 7945,
South Africa}
\affiliation{Department of Mathematics, University of Stellenbosch, Stellenbosch 7602, South Africa}
\author{Arrykrishna Mootoovaloo}
\affiliation{Department of Mathematics and Applied Mathematics, University of Cape Town,
Rondebosch, Cape Town, 7700, South Africa}
\affiliation{African Institute for Mathematical Sciences,
6–8 Melrose Road,
Muizenberg 7945,
South Africa}
\affiliation{South African Astronomical Observatory, Observatory Road, Observatory, Cape Town, 7935,South Africa}
\author{Hans Eggers}
\affiliation{Department of Physics,
Stellenbosch University,
P/Bag X1,
7602 Matieland, South Africa }
\affiliation{National Institute for Theoretical Physics, Stellenbosch, South Africa}
\author{Zafiirah Hosenie}
\affiliation{Centre for Space Research, North-West University, Potchefstroom 2520, South Africa}
\affiliation{African Institute for Mathematical Sciences,
6–8 Melrose Road,
Muizenberg 7945,
South Africa}
\affiliation{South African Astronomical Observatory, Observatory Road, Observatory, Cape Town, 7935,South Africa}
\author{Steve Kroon}
\affiliation{CSIR-SU Centre for AI Research,
Computer Science Division,
Stellenbosch University,
P/Bag X1,
7602 Matieland, South Africa}
\author{Elena Sellentin}
\affiliation{Imperial Centre for Inference and Cosmology (ICIC), Imperial College, Blackett Laboratory, Prince Consort Road, London SW7 2AZ, U.K.}
\affiliation{D\' epartement de Physique Th\' eorique,
Universit\' e de Gen\` eve,
Quai Ernest-Ansermet 24
CH-1211 Gen\` eve, Switzerland}

\date{\today}

\maketitle

\section{Abstract}
In this paper, we present a method for computing the marginal likelihood, also known as the model likelihood or Bayesian evidence, from Markov Chain Monte Carlo (MCMC), or other sampled posterior distributions. In order to do this, one needs to be able to estimate the density of points in parameter space, and this can be challenging in high numbers of dimensions.  Here we present a Bayesian analysis, where we obtain the posterior for the marginal likelihood, using $k$th nearest-neighbour distances in parameter space, using the Mahalanobis distance metric, under the assumption that the points in the chain (thinned if required) are independent.  We generalise the algorithm to apply to importance-sampled chains, where each point is assigned a weight.  We illustrate this with an idealised posterior of known form with an analytic marginal likelihood, and show that for chains of length $\sim 10^5$ points, the technique is effective for parameter spaces with up to $\sim 20$ dimensions.  We also argue that $k=1$ is the optimal choice, and discuss failure modes for the algorithm.  In a companion paper (Heavens et al. 2017) we apply the technique to the main MCMC chains from the 2015 \Planck\ analysis of cosmic background radiation data, to infer that quantitatively the simplest 6-parameter flat $\Lambda$CDM standard model of cosmology is preferred over all extensions considered.

\section{Introduction}

The marginal likelihood is an important quantity in Bayesian analysis, as it allows model comparison; in conjunction with prior information on the models, it provides posterior probabilities of competing models, given a dataset.  It may be challenging to compute, as it formally involves an integral over what may be a high-dimensional model parameter space.  Typically such parameter spaces are explored for parameter inference with sampling methods such as Monte Carlo Markov Chains (MCMC).  It would be of value if such chains could also be used to infer the marginal likelihood.  This can be done by exploiting the fact that asymptotically the density of MCMC points is proportional to the target density, which is typically the likelihood, or the likelihood multiplied by the prior, which together form an unnormalised posterior density. If we are able to determine the constant of proportionality, then we can infer the marginal likelihood from the MCMC chain.  While inference of densities from point samples in high dimensions suffers from the
curse of dimensionality, we show here that a Bayesian method  using the ($k$th) nearest neighbour distance, with a suitably-defined metric, can be effective for determining this unknown constant.  This makes use of the fact that every point in the MCMC chain provides a likelihood of the unknown constant, so the posterior from the whole chain can be precise and accurate.  

In Bayesian parameter inference, we seek the posterior distribution of model parameters, represented by a vector $\vtheta$, given a data set $\vx$, any prior information (implicit) and a model $M$:
\be
p(\vtheta|\vx, M) = \frac{p(\vx|\vtheta, M)\,\pi(\vtheta|M)}{p(\vx|M)}
\ee
where $p\left(\vx | \vtheta, M\right)$ is the likelihood,
which in Bayesian analysis is regarded as a function of $\vtheta$
with the dataset $\vx$ being fixed, and $\pi\left(\vtheta | M\right)$
is the prior.
The evidence $p(\vx|M)$ normalises the right hand side such that the posterior is a genuine probability density for $\vtheta$, and
is the integral over the unnormalised posterior $\tilde{p}\left(\vtheta | \vx, M\right) \equiv p\left(\vx |  \vtheta,M\right)\pi\left(\vtheta | M\right)$:
\be
E \equiv p(\vx|M) = \int d\vtheta\, p(\vx|\vtheta,M)\,\pi(\vtheta|M),
\label{eq: Marginal Equation}
\ee
so is often referred to as the marginal likelihood.  It is of fundamental importance in Bayesian model comparison because
it plays a role in the posterior model probability.  Bayes\textquoteright{} theorem applied to
models gives the relative posterior probability of competing
models as
\be
\frac{p(M_1|\vx)}{p(M_2|\vx)} = \frac{\pi(M_1)}{\pi(M_2)}\,\frac{p(\vx| M_1)}{p(\vx|M_2)}
\ee
which is the ratio of the model priors multiplied by the Bayes factor,
or ratio of the marginal likelihoods \citep{Jeffreys}. In general,
eq. \ref{eq: Marginal Equation} may be a difficult integral to
compute, possibly involving a high-dimensional integral, and
an integrand which may be expensive. Various approximate
methods and techniques have been suggested to evaluate the marginal likelihood from a set of samples drawn from the target distribution (e.g., the posterior), using for example MCMC techniques \citep{Metropolis,Hastings,Geman}.  We summarise some of these here. \citet{Newton}
employed the Harmonic mean estimator which is simple to implement
in MCMC methods; however it outputs high variability in the estimator, which 
can even have infinite variance.  Improvements were made by \citet{PetrisTardella}.
\citet{Chib} utilized the output from a Gibbs sampler for $p\left(\vtheta | \vx ,M\right)$
to evaluate the marginal likelihood $p\left(\vx | M\right)$.
However, the algorithm depends on a block updating technique
for $\vtheta$ which is not always possible. Therefore, \citet{jeliakov} expanded this work to estimate the marginal likelihood
by using output from a Metropolis-Hastings method \citep{Metropolis,Hastings,Tierney,Greenberg}
for the posterior $p\left(\vtheta|\vx ,M\right)$. The method
that \citet{Chib} and \citet{jeliakov} applied can be very precise
provided that enough MCMC samples are drawn, yet needs
to be adapted for each specific case. In addition, \citet{Chib2005}
described an algorithm for approximating the marginal likelihood in the case of comparing
models via Bayes factors, using accept-reject Metropolis-Hastings
(ARMH) methods \citep{Tierney,Greenberg}.

\citet{marin} proposed Approximate Bayesian Computation (ABC),
which is more complex than MCMC, but outputs cruder estimates, and
\citet{Jordan} utilised variational methods that yield lower
bounds for the marginal likelihood. Variational methods have also been used by
\citet{McGrory} for model comparison in mixture models. Further, \citet{Friel}
implemented the power posterior methods and \citet{Neal} used the
combination of some concepts from simulated annealing and importance
sampling to compute the marginal likelihood. \citet{Meng} implemented
a bridge function using some ideas from bridge sampling to calculate
the Bayes factor by combining the two posterior distributions. The
popular Integrated Nested Laplace approximation (INLA) applied by
\citet{Rue} performs the estimation of the marginal likelihood within the class of
latent Gaussian structures. Also, in the case of non-nested sampling,
work has been done on the direct approximation of the marginal likelihood \citep{Chib,Gelfand}
and especially in nested models, approximate ratios of 
marginal likelihoods have been computed \citep{Chen,DiCiccio,Meng,Verdinelli}.   Finally, there are methods \cite{multinest,polychord} that are designed explicitly to provide estimates of the marginal likelihood.  

In this paper, we propose a method that adds value to chains sampled for parameter inference, by also using them to calculate the marginal likelihood for the model considered

The layout of the paper is as follows.  In section \ref{method} we set out the method, based on local density estimation of the chain in parameter space.  In subsection \ref{BayesML} we develop a Bayesian method to compute the posterior for the marginal likelihood; in subsection \ref{importance} we generalise the result to importance-sampled chains, and in section \ref{PW} we improve the algorithm by using the Mahalanobis distance instead of a naive distance metric. We study a test case in section \ref{results}, and finally discuss the limitations and applicability of the algorithm in section \ref{conclusions}.

\section{Method}
\label{method}

Properly-designed MCMC samples have the property that the expectation value of their number density, $n(\vtheta|\vx, M)$ is proportional to the parameter-space target density, in this case the unnormalised posterior (defined to be the likelihood times a properly-normalised prior):
\[
\post = a\,n(\vtheta|\vx, M)
\]
for some constant $a$ (which will scale with the total number of samples $N$). If we are able to determine $a$, then the marginal likelihood (here denoted $E$) follows immediately from the sampled $n$, which is a sum of Dirac delta functions:
\begin{eqnarray}
E = p(\vx|M) &=& a\int d\vtheta \,n(\vtheta|\vx, M) \nonumber\\
&=& a\int d\vtheta \,\sum_{\alpha=1}^N\,\delta(\vtheta-\vtheta_\alpha)  = a N
\label{EaN}
\end{eqnarray}
where $N$ is the length of the chain.  Alternatively, we note that $n=N p(\vtheta|\vx,M) = N \tilde p/E$ from which $E=N\tilde p/n \equiv aN$.  This is modified slightly for importance-sampled or weighted chains; we consider these in section \ref{importance}. Chib (1995) proposed that one might be able to determine the posterior density at one point, and use this to determine $a$.  Here we propose an alternative approach that uses the entire chain.  

\subsection{Bayesian determination of the marginal likelihood}
\label{BayesML}

Inferring the density from a set of point samples is not necessarily a trivial task, especially in high dimensions.  We assume that the points in the chain are independent, which may require some thinning of a chain, depending on the sampling method.  This would almost certainly be the case for Metropolis-Hastings samples, less so for Gibbs or Hybrid Monte Carlo.

Many methods of density estimation exist. In this paper, we investigate the use of the $k$th nearest neighbour distance to infer $a$, since it is dependent on the local density of points and hence on $\post$.  We develop the formalism for general $k$, and discuss the benefits of different choices for $k$, concluding that $k=1$ is preferred.  Nearest neighbour distances have the advantage that each point gives a likelihood for $a$,  and the very many contributions to the likelihood from the chain can be combined in a Bayesian way to obtain a posterior for the marginal likelihood.

Consider a sample with constant expected number density $n$. If the parameter space has dimension $m$, then the pdf of the distance $D_k$ to the $k$th nearest sample point from an arbitrary point is specified by the requirements that there are exactly $k-1$ sample points within $D_k$ of the point, and precisely one within a range $(D_k, D_k+dD_k)$. Poisson sampling gives
\begin{eqnarray}
p(D_k | n)dD_k &=& \frac{[n V_m(D_k)]^{k-1} e^{-n V_m(D_k)}}{(k-1)!} \nonumber\\
&\times& n \, dV_m(D_k) e^{-n \,dV_m(D_k)}
\end{eqnarray}
where $V_m(D_k) = \pi^{m/2} D^m_k/\Gamma(1+m/2)$ is the volume of the $m-$ball of radius $D_k$.  Thus in the limit $dV_m\rightarrow 0$,
\[
p(D_k | n) = \frac{n^k V^{k-1}_m e^{-n V_m}}{(k-1)!}  \frac{dV_m}{dD_k}.
\]
If we assume that the sampling density is high enough that $n$ can be considered uniform over a volume $\gg V_m$ for typical $k$th nearest neighbour distances, then we can use Bayes' theorem to obtain the likelihood of $a$ from a single sampled point in the chain, labelled by $\alpha$:
\be
p(a | D_{k,\alpha},M) \propto p(D_{k,\alpha} | a) \pi(a) \propto p\left(D_{k,\alpha} | n=\frac{\tilde p_\alpha}{a}\right) \pi(a) 
\label{pa}
\ee
where $\pi(a)$ is the prior on $a$, $\tilde p_\alpha$ is the sample value of the unnormalised posterior, and we have suppressed the dependence on the model for clarity.  Note that we require that the chain has recorded the target distribution $\tilde p$ along with the parameter values. Assuming independence of samples (strictly, we assume that the $k$th nearest-neighbour distances are independent, for a given $k$) the posterior for $a$ from the entire chain, represented as a vector of $k$th nearest neighbours $\vD$, is
\be
p(a | \vD) \propto \left[\prod_{\alpha=1}^N p\left (D_{k,\alpha} | n=\frac{\tilde p_\alpha}{a}\right ) \right] \pi(a).
\ee
With the assumed Poisson sampling 
\be
\ln p(a | \vD) = {\rm cst.} - N k\ln a -\frac{1}{a}\sum_{\alpha=1}^N V_m(D_{k,\alpha}) \tilde p_\alpha + \ln \pi(a).
\ee
Hence we find the posterior for the evidence, or marginal likelihood $E = aN$:
\be
\ln p(E| \vD, M) =  {\rm cst.} - N k\ln E -\frac{N}{E}\sum_{\alpha=1}^N V_m(D_{k,\alpha}) \tilde p_\alpha + \ln \pi(E).
\ee
Since $a$ (and hence $E$) is a scaling parameter, we choose a Jeffreys prior $\pi(E) \propto 1/E$ to obtain the posterior
\be
\ln p(E| \vD, M) =  {\rm cst.} - (N k+1)\ln E -\frac{N}{E}\sum_{\alpha=1}^N V_m(D_{k,\alpha}) \tilde p_\alpha.
\label{postE}
\ee
The maximum posterior value of $E$ is
\be
E_{\rm MAP} = \frac{N\sum_{\alpha=1}^N V_m(D_{k,\alpha})\tilde p_\alpha}{N k+1}.
\ee
and the posterior fractional variance is approximately (from the second derivative at the peak)
\be
\frac{\sigma_E^2}{E^2} = \frac{1}{Nk+1}.
\ee
The approximation of independence is likely to be good for a suitably thinned chain. In many (but not all) cases, if A is the nearest neighbour of B, then B is the nearest neighbour of A, so we expect this variance to be underestimated by a factor $\sim 2$.

We anticipate that this technique will work in small numbers of dimensions, but fail (for fixed $N$) as the dimensionality increases, when the $k$th nearest neighbour is typically at a distance over which the target distribution is not well-approximated by a constant.

\subsection{Importance-sampled case}
\label{importance}

We repeat the calculation for the case when the chain samples from a distribution $q(\vtheta)$ that is different from $\tilde p$:
\be
q(\vtheta) = \frac{\tilde p(\vtheta|\vx,M)}{w(\vtheta)}
\ee
for some weight function, and each sample point has a weight $w_\alpha= w(\vtheta_\alpha)$.

Equation (\ref{EaN}) is modified, since now $q = a n$, so 
\be
n = \frac{q}{a} = \frac{\tilde p}{w a},
\ee
so
\be
E = \int \tilde p(\vtheta|\vx,M)\,d\vtheta = a\int w(\vtheta)\sum_{\alpha=1}^N \delta(\vtheta-\vtheta_\alpha) \,d\vtheta = a W,
\ee
where $W\equiv \sum_{\alpha=1}^{N} w_\alpha$ is the sum of weights.  Hence
\be
\ln p(a | \vD,M) = {\rm cst.}+ \sum_{\alpha=1}^N \ln p\left (D_{k,\alpha} | n=\frac{\tilde p_\alpha}{w a}\right ) +\ln\pi(a).
\ee
With the assumed Poisson sampling 
\be
\ln p(a | \vD,M) = {\rm cst.} - N k\ln a -\frac{1}{a}\sum_{\alpha=1}^N \frac{V_m(D_{k,\alpha}) \tilde p_\alpha}{w_\alpha} + \ln \pi(a).
\ee
Hence we find the posterior for the marginal likelihood, assuming a Jeffreys prior as before:
\be
\ln p(E| \vD, M) =  {\rm cst.} - (N k+1)\ln E -\frac{W}{E}\sum_{\alpha=1}^N \frac{V_m(D_{k,\alpha}) \tilde p_\alpha}{w_\alpha}.
\label{postEW}
\ee
The maximum posterior value of $E$ is
\be
E_{\rm MAP} = \frac{W}{N k+1}\,{\sum_{\alpha=1}^N \frac{V_m(D_{k,\alpha})\tilde p_\alpha}{w_\alpha}}.
\ee
and the estimate of the posterior fractional variance is unchanged.  
Equation (\ref{postEW}) for the posterior for the marginal likelihood, or Bayesian Evidence, is the principal result of this paper.

\subsection{Pre-whitening}
\label{PW}

Since the parameters in a model may have very different units, there is no guarantee that the variances of the posterior will be comparable in each dimension, and in general there will also be correlations.  It is necessary to define a dimensionless nearest-neighbour distance, and in order to treat all parameters on the same footing, we diagonalise the parameter space and then rescale to make the variances equal, before measuring nearest-neighbour distances using a Euclidean metric.   This is effectively a pre-whitening step, and is equivalent to using the Mahalanobis distance in the original space\cite{mahalanobis}, which uses the covariance matrix to define a metric tensor for the original parameter space.  It proves to be very effective for unimodal likelihoods, and proceeds as follows.

From the chain, we compute the covariance matrix $C_{ij} \equiv \langle (\vtheta - \bar\vtheta)_i (\vtheta - \bar\vtheta)_j\rangle$
and then we diagonalise in the usual way, and finally rescale the new linear combinations of parameters by the square root of the eigenvalues of $C$, such that their covariance matrix is the identity.  The transformation of number density involves division by the Jacobian $J=\sqrt{\det(C)}$.

\section{Results}
\label{results}

We illustrate the method with a simple case that has an analytic solution.  The model is that $m$-dimensional vectors are drawn independently from a multivariate gaussian, $\vx \sim {\mathcal{N}}(\vmu,\Sigma)$, where the (fixed) covariance matrix $\Sigma$ is arbitrary (generated to be positive-definite by generating a random matrix $\A$ and forming $\Sigma=\A^T\A$).   The parameters in the model are the $m$ expectation values, $\vmu$.  In this case, the likelihood for a set of $n$ vectors (represented by the data vector $\vx_i$; $i=1\ldots n$) is
\begin{eqnarray}
p(\vx | \vmu,M) &=& \prod_{i=1}^n  \frac{1}{\sqrt{|2\pi \Sigma|}}\exp\left[-\frac{1}{2}(\vmu-{\vx_i})^T\Sigma^{-1}(\vmu-{\vx_i})\right] \nonumber\\
& =  & \frac{1}{|{2\pi \Sigma|^{n/2}}} \exp\left[-\frac{n}{2} (\vmu-\bar\vx)^T\Sigma^{-1}(\vmu-\bar\vx)-\frac{1}{2}\sum_{i=1}^n (\vx_i-\bar\vx)^T\Sigma^{-1}(\vx_i-\bar\vx)\right]\nonumber\\
\end{eqnarray}
and $\bar\vx=n^{-1}\sum_{i=1}^n \vx_i$ is the sample mean. 

\begin{figure}
\includegraphics[width=8cm]{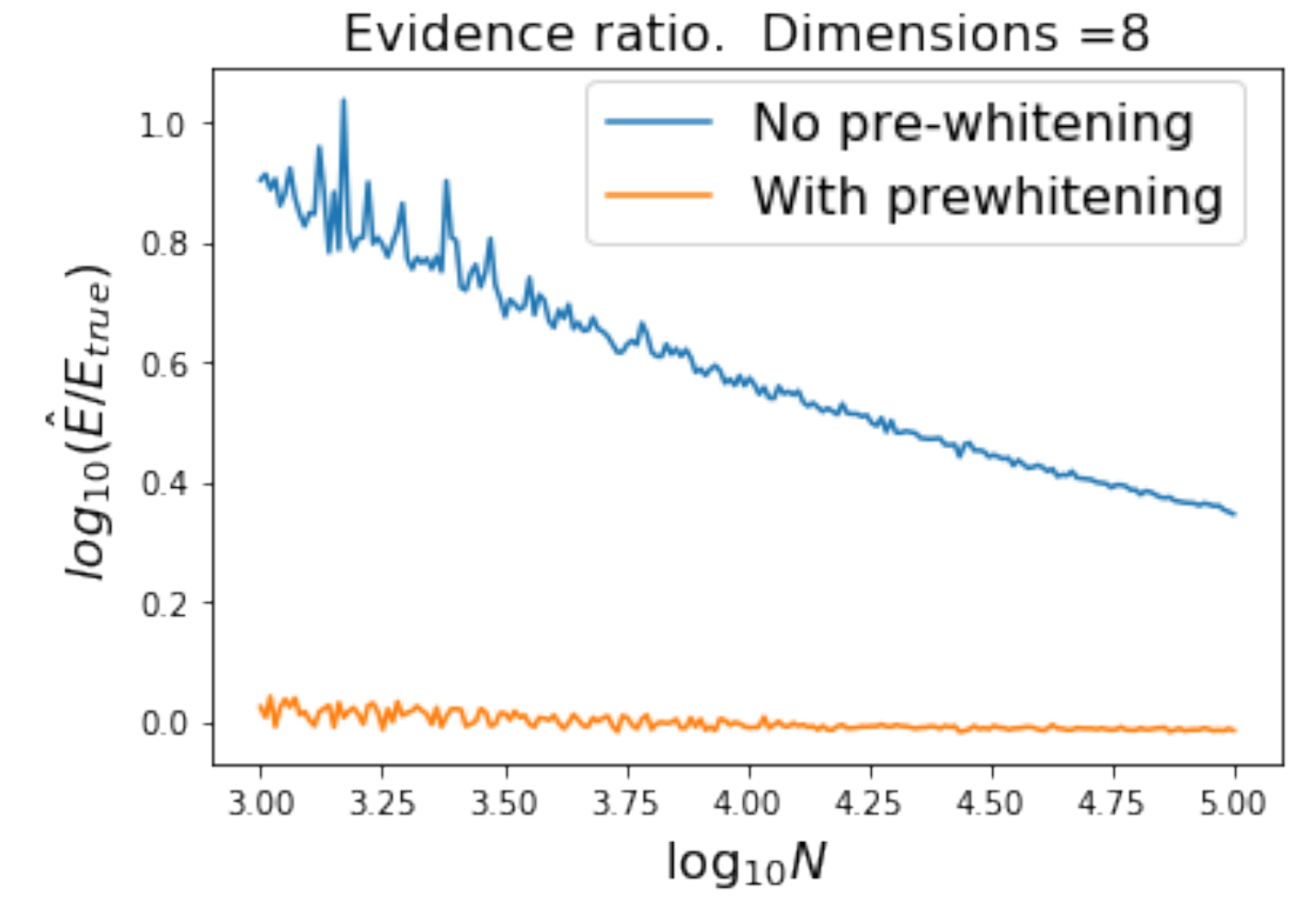}
\caption{Log of the MAP marginal likelihood estimate relative to the analytic solution, for 8 dimensions and a random multivariate gaussian posterior, with $k=1$.  Orange curve shows the algorithm with pre-whitening, blue is without.}
\label{Fig1}
\end{figure}

\begin{figure}
\includegraphics[width=8cm]{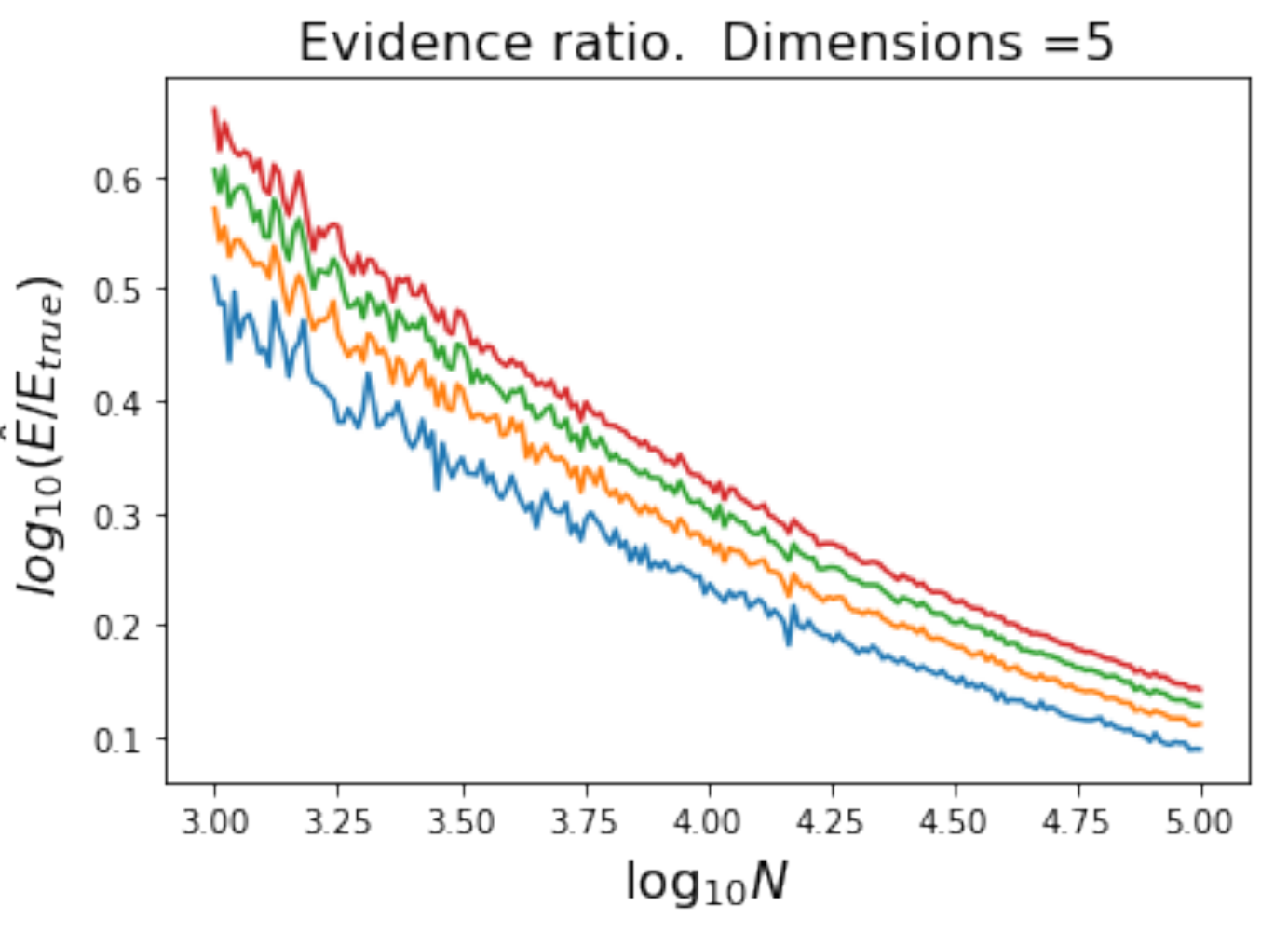}
\caption{Similar to Fig. \ref{Fig1}, but showing $k=1$ to $4$ (bottom to top), without prewhitening.  $k=1$ is preferred.}
\label{Fig2}
\end{figure}

\begin{figure}
\includegraphics[width=8cm]{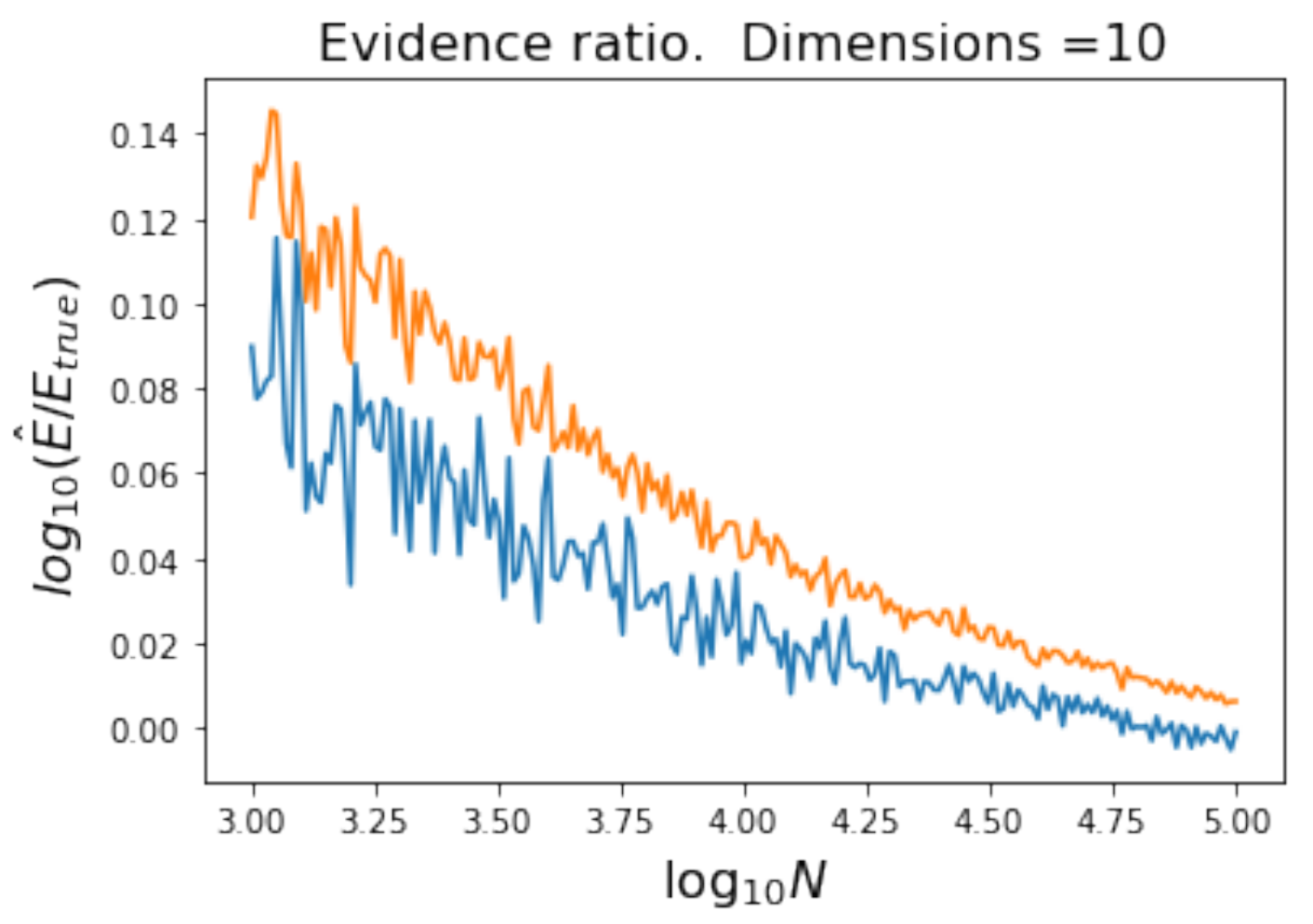}
\caption{Marginal likelihood compared to analytic result, for dimension $m=10$ and prewhitened algorithm; $k=1$ (lower) and $k=4$.  $k=1$ is noisier, as expected, but has higher accuracy and is preferred.}
\label{Fig3}
\end{figure}

If we assume a uniform prior for each component of $\vmu$, with widths $\Delta\mu_j$ sufficiently large that the likelihood is negligible outside the range, the prior is $\pi_0 = \prod_{j=1}^m (\Delta\mu_j)^{-1}$ and the marginal likelihood can be integrated by extending the integrals to infinity, yielding a marginal likelihood
\be
E = \frac{\pi_0}{n^{mn/2}}\exp\left[-\frac{1}{2}\sum_{i=1}^n (\vx_i-\bar\vx)^T\Sigma^{-1}(\vx_i-\bar\vx)\right].
\ee
In Fig.\ref{Fig1} we plot the ratio of the maximum a posteriori (MAP) value of the evidence $\hat E$ to the analytic evidence, for an 8-dimensional random gaussian, with 1000 data points, as a function of the logarithm of the number of samples.  $\pi_0$ cancels in the ratio if points are sampled from the likelihood. MAP estimates are shown with and without pre-whitening, for $k=1$.  Fig.\ref{Fig2} shows $m=5$, and we show the dependence on $k$ with $k=1\ldots 4$ without prewhitening.  We see the best results for the MAP estimate
with k = 1, but none of these results is very accurate.   In Fig. \ref{Fig3} we show the results  with prewhitening for $m=10$, for $k=1$ and $k=4$, where the same preference for $k=1$ is seen, but the accuracy is much better.  In 20 dimensions, the accuracy for random gaussian distributions is about a factor of two with $10^5$ MCMC points.  Given that evidence calculations have some prior sensitivity, this level of accuracy may still be useful in model comparison.
 
\section{Discussion}
\label{conclusions}

In this paper, we have used the $k$th nearest neighbour distances to calculate the marginal likelihood, or Bayesian evidence, from Monte Carlo Markov Chains.   MCMC methods are designed to produce samples of the target distribution, with an expected number density proportional to the target density. The constant of proportionality is required in order to compute the marginal likelihood.  Since the nearest neighbour distances are dependent on the local number density of points in parameter space, they contain information on the unknown constant, and although the posterior for the density is high from a single point, typical chains have many points, and the information can be combined in a Bayesian way to provide a posterior probability distribution function for the marginal likelihood. We have generalised the method to treat importance-sampled chains.  Other density estimation methods may also be effective for this problem; the $k$th nearest neighbour method is convenient in that it yields very naturally to a Bayesian analysis.  

We have shown that in the case of random multivariate gaussian target distributions that the method fails badly in more than a few dimensions if a naive Euclidean metric is used.  However, by using the Mahalanobis distance, based on the covariance matrix of the MCMC chain, the method is accurate for this problem at percent level up to 10 dimensions (with chains of length $10^5$), and to a factor of $\sim 2$ in 20 dimensions.  An equivalent procedure to the Mahalanobis distance is to pre-whiten the chain, by rotating the parameter space to the principal axes of the covariance matrix, thus diagonalising it, and then rescaling the new parameter combinations so that each axis has unit variance. In this system, a Euclidean metric is used, and the marginal likelihood computed with due regard for the Jacobian of the transformation.

Some caveats are in order.  The algorithm assumes that the points are independent, which will not be strictly true for typical MCMC chains. A correlation analysis should be performed, and the chain thinned if required such that the points are at least only weakly correlated. Secondly, a failure mode is if the target distribution is not well-approximated by a constant over the typical nearest-neighbour distance.  This is the origin of the inaccuracies seen in the figures in this paper.  These grow with dimension and with decreasing sample size, with significant (0.1 dex) errors occurring when $(\alpha_m N/V)^{-1/m}>0.5$, where $V_m \equiv \alpha_m D^m$ and $V$ is the volume of the target, here equal to unity. The inaccuracy also increases with $k$, so the nearest-neighbour distance itself ($k=1$) is optimal.  For a complicated target distribution that is not necessarily mono-modal, the method may fail as there may be fine structure in the target.  In such cases, there may be some scope for improvement by using a local Mahalanobis distance, rather than one based on the global properties of the chain.  If one has an efficient method of computing the expected curvature matrix of the target distribution, via for example the Fisher matrix, then this could be a feasible extension to the method. 

The code for computing the marginal likelihood from sampled target distributions is available on Github at https://github.com/yabebalFantaye/MCEvidence.

\acknowledgments
We thank Andrew Jaffe, Michiel De Kock, Bruce Bassett , Roberto Trotta and David van Dyk for useful discussions. The idea for this paper was advanced at the 2016 Bayes School and Workshop in November 2016 in Stellensbosch, South Africa, funded by NITHeP. YF is supported by the Robert Bosch Stiftung. 

\bibliography{ref}

\end{document}